\documentclass{optica-article}

\journal{opticajournal} 

\articletype{Research Article}

\usepackage{lineno}

\begin{document}

\title{Fault-Tolerant Four-Dimensional Constellation for Coherent Optical Transmission Systems} 

\author{Jingtian Liu\authormark{$\ast$}, \'Elie Awwad, Yves Jaou\"{e}n}

\address{Communication and Electronics department, LTCI, Télécom Paris, Institut Polytechnique de Paris, 19 place Marguerite Perey, 91120 Palaiseau, France\\}

\email{\authormark{*}jingtian.liu@telecom-paris.fr} 



\begin{abstract*}
We propose a 4-dimensional 2-ary amplitude ring-switched modulation format with 64 symbols, which is denoted as 4D-2A-RS64 encoded over two polarization tributaries to improve the transmission performance over long-haul optical fibers in the presence of the non-linear Kerr effect. At a spectral efficiency of 6 bits per 4D, simulation results show that this format outperforms the polarization division multiplexed (PDM) 8QAM-star modulation as well as the 4D-2A-8PSK over links without inline dispersion management. 
We evaluate the performance for a WDM transmission of $11\times90~\mathrm{Gbaud}$ channels over a multi-span SSMF link. For an achievable information rate of $4.8\mathrm{bit/s/Hz}$, the maximum transmission distance is improved by $10.6\%$ (400 km) and $4\%$ (160 km) compared to PDM-8QAM-star and 4D-2A-8PSK respectively. The achieved gains are composed of a linear part and a non-linear part, respectively from the improved Euclidean-distance distribution and the constant power property of the 4D modulation. The geometric shaping of the proposed scheme is easy to implement and is robust to Mach-Zehnder modulator (MZM) imbalances and quantization errors stemming from the finite digital-to-analog converter (DAC) resolution. This robustness is compared to the one of other geometric-shaped non-linearity tolerant 4D schemes such as the 4D-2A-8PSK and the 4D-64PRS that can be both outperformed by our scheme in severe conditions.

\end{abstract*}


\section{Introduction}
In recent years, several studies have shown that multi-dimensional modulation formats can improve the performance of coherent transmission systems as they offer higher Euclidean distances, better spectral efficiency (SE) granularity, and a potential to mitigate nonlinear fiber effects compared to conventional formats such as polarization division multiplexed 2D formats like PDM quadrature phase shift keying (PDM-QPSK) or PDM $M$-QAM (quadrature and amplitude modulation with $M$ 2D symbols)~\cite{ref8,welti1974digital}. For instance, in~\cite{ref9}, the mitigation of cross-polarization modulation (XpolM) was realized by an 8D-QPSK using the polarization balanced concept with an SE of 2 bits over two polarization states and one time slot (4D). The experimental demonstration showed a $1$~dB improvement in net system margin relative to PDM Binary-Phase-Shift-Keying (PDM-BPSK) over dispersion-managed (DM) systems. Later on, the same concept was used in~\cite{ref10} to design modulation formats with a higher SE of 2.5 and 3.5 bits per 4D symbol respectively. However, their non-linear gains in DM systems vanish over dispersion-unmanaged (DUM) systems, because a large accumulated dispersion destroys their 8D structure. In this work, our interest is in achieving performance gains over DUM systems using low-complexity multi-dimensional formats and aiming at higher SE values.

Over the DUM system, for a higher SE of 6 bits per 4D symbol, several 4D modulation formats such as 4D-64 Set Partitioned (SP)-12QAM~\cite{ref11}, Honey Comb (HC) grid~\cite{ref22} and others~\cite{ref25,ref26,ref23} were designed by increasing the Euclidean distances and enhancing the non-linearity tolerance through the minimization of the peak-to-average-ratio (PAPR) of the constellation which leads to lower non-linear interference (NLI) generated by both intra-channel and inter-channel effects~\cite{ref33}.  


Later on, Kojima et al. in~\cite{ref2,ref17} proposed the 4D-2A-8PSK achieving higher gains, in both linear and nonlinear regimes, than PDM-8QAM-star, PDM-8PSK, and circular PDM-8QAM over DM systems. The gains stem from a larger Euclidean distance, a  4D constant modulus constraint, and Gray labeling. The Euclidean distance gain comes from the nesting of two 8PSK constellations of different amplitudes followed by set partitioning. The complementary amplitudes of the two polarization tributaries ensure a constant energy property at each time slot, thus eliminating energy variations and reducing nonlinear effects. The Gray labeling also brings an improved achievable information rate (AIR) compared to PDM-8QAM-star which is not Gray labeled. The authors were able to achieve flexible SE values from 5 to 7 bit/4D through tailored mapping using Boolean equations. The ring ratios of the 4D-2A-8PSK are computed based on an optimization of the AIR at different signal-to-noise ratios (SNR)~\cite{ref32}. However, its optimized ring ratio varies between $0.6$ and $1$ for SNR values between $5$ and $12$~dB, which shows that it is very sensitive to changes in the ring ratio.  The performance of 4D-2A-8PSK was not explored over DUM systems.

Then, based on the 4D-2A-8PSK, a 4D-64PRS constellation (PRS standing for polarization ring switching) was proposed in~\cite{ref14} by a new mapping technique named `Orthant-symmetric labeled constellation', a method that maps the possible combinations in quadrants separately to binary labels to facilitate the upcoming set partitioning. The 4D-64PRS has a slightly lower minimum squared Euclidean distance (MSED) than 4D-2A-8PSK but also has a lower kissing number, i.e. number of symbol pairs at MSED, which yields gains of up to 0.7 dB in SNR in the AWGN channel compared with PDM-8QAM-star. It also has a better performance in the presence of non-linear effects because of its 4D constant modulus constraint. AIR-based optimization combined with geometric shaping is used to find the optimum ring ratio and angle for SNR levels between $3$ and $13$~dB. The optimized ring ratio varies in the range of $0.53-0.61$ and the relative rotation angle between the two rings is fine-tuned between $23.4$ to $27.2$ degrees. Both methods combine geometric shaping~\cite{ref28,ref29} and AIR-based optimization~\cite{ref30,ref18} to find the best ring ratio and constellation point locations that maximize the AIR at a given SNR value. More recently, new geometric shaping schemes achieving higher spectral efficiency were also investigated in~\cite{sillekens2022high,goossens2022introducing}. The fine geometric optimizations increase the constraints on the generation of the designed constellation (DAC resolution, stability, and accuracy of optical modulators,...). Moreover, their performance fluctuates considerably with the design parameters.

Apart from geometric-shaped multi-dimensional designs, probabilistic constellation shaping (PCS) based schemes~\cite{wu2021temporal,wu2022list,neskorniuk2023memory} were also proposed to offer both linear and non-linear gains; however, they are not considered in this work in which we focus on transmissions that do not use distribution matching to shape and un-shape the probabilities of the transmitted symbols at the transmitter and receiver side respectively. Leveraging PCS and multi-dimensional modulations for mitigation of Kerr-induced non-linear interference is left for future investigation. In this paper, we propose a new 4-dimensional structure based on an 8-PSK modulation surrounding a QPSK on each polarization, with the outer ring having twice the value of the inner ring. The Peak-to-Average Power Ratio (PAPR) of the transmitted constellation is minimized to one by ring switching over the two polarization tributaries. The new modulation has an SE of 6 bits per 4D. Through numerical simulations, we demonstrate the advantage in non-linearity tolerance compared to PDM-8QAM format and other state-of-the-art 4D formats especially when severe transmitter imperfections are considered. 

Achievable information rates (AIRs) are information-theoretic metrics related to the amount of reliable information that can be transmitted over a given channel. AIRs usually include symbol-wise mutual information (MI) and generalized mutual information (GMI), also called bit-wise MI~\cite{ref18,ref19}. While MI is a measure of the maximum amount of transmittable information between the received and transmitted symbols, GMI takes into account the effect of the bit mapping and de-mapping. Both metrics have been extensively used to design modulation formats and to predict the ultimate performance of discrete modulation formats with forward error correction (FEC) codes~\cite{ref20,ref21}. In this paper, we choose GMI as the performance metric and compute it through numerical simulations as the achievable rate of a bit-wise decoder defined as Eq.~(16) in~\cite{ref31}.

The paper is organized as follows. In section~\ref{Principle}, we introduce the structure of the designed 4D-2A-RS64 and then analyze in section~\ref{Tolerance} its robustness to changes induced by I-Q gain and quadrature imbalances and to DAC quantization at the transmitter side. In section~\ref{Performance evaluation}, numerical results for our proposed 4D modulation and three other modulation formats are shown, over an AWGN channel and then over a non-linear WDM multi-span transmission. Finally, we conclude by highlighting the benefits of the novel modulation 4D scheme.

\section{Principle}
\label{Principle}

\begin{figure}[t]
    \centering
    \includegraphics[width=1 \linewidth]{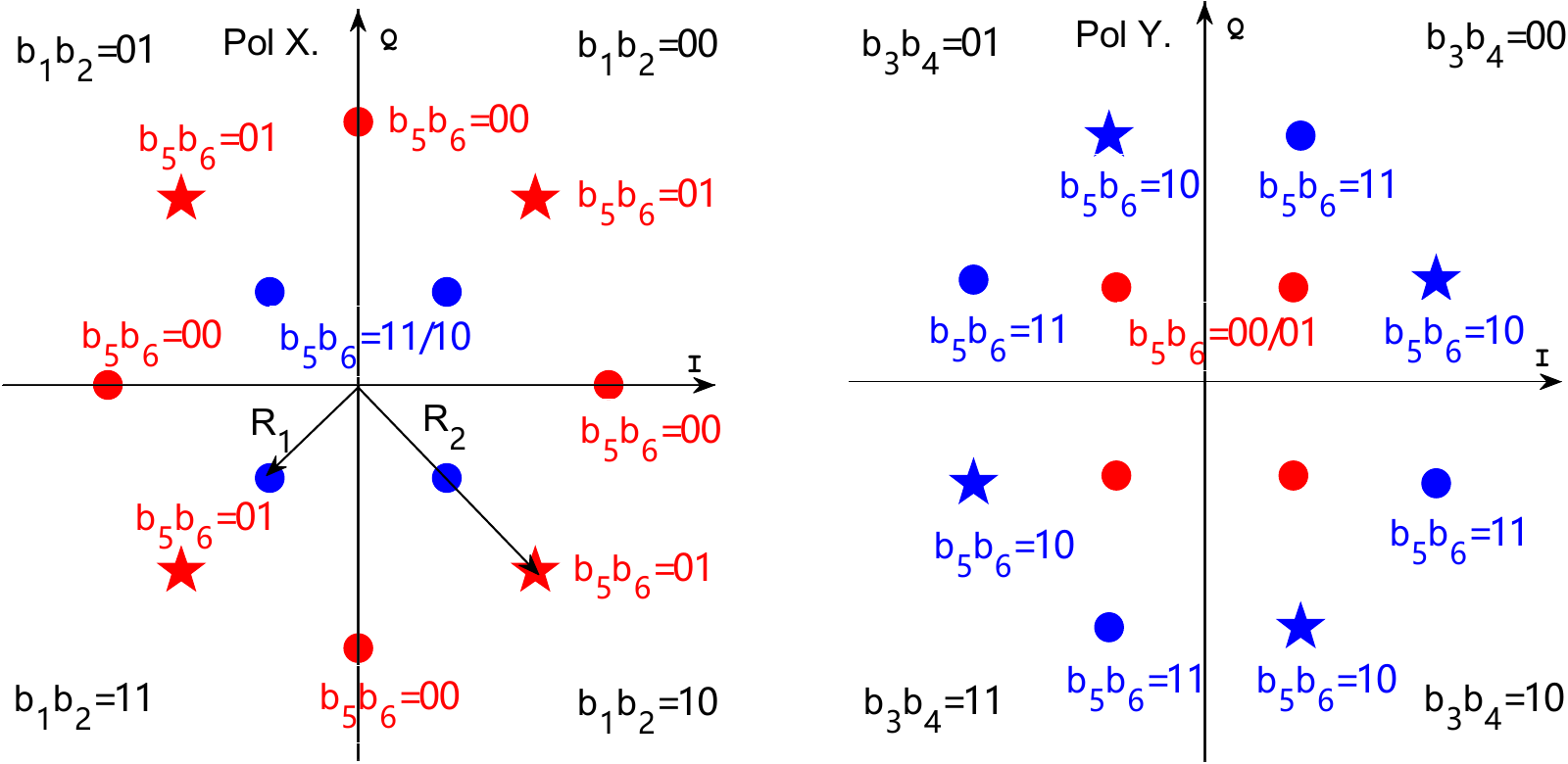}
    \caption{2D-projections of the designed 4D-2A-RS64 modulation and associated bit-to-symbol mapping with $R_2=2R_1$. Red (resp. blue) circle markers and star markers on $R_2$ have the same labeled $\left[b_5,b_6\right]$ bits.}
    \label{fig:figure1}
\end{figure}

In multi-span transmission systems, non-linear interference (NLI) arising from the Kerr effect, and accumulated amplified spontaneous emission noise from EDFAs dominate the AIR performance of a given modulation format. Fewer temporal power variations help in reducing the former and a larger minimum squared Euclidean distance (MSED) of the constellation increases the performance in the presence of the latter. Hence, we conceive our constellation in this spirit. Fig.~\ref{fig:figure1} shows the construction and binary labeling of the proposed scheme, named four-dimensional two-amplitude ring-switched constellation with 64 symbols: 4D-2A-RS64. Two PSK modulations are considered: a QPSK with a radius $R_1$ and an 8-PSK with a radius $R_2=2R_1$. The 8-PSK on Y polarization (Pol Y) has a phase rotation of $\frac{\pi}{8}$ with respect to the QPSK on X polarization (Pol X) to optimize the distribution of the Euclidean distances between 4D symbol pairs as will be shown in Section~\ref{Performance evaluation}.A. Then, a 4D symbol, defined as two I-Q symbols multiplexed on two orthogonal polarization tributaries over one time slot, exists only when the two 2D symbols are colored the same in Fig.~\ref{fig:figure1}, which means that two 2D symbols from different rings $R_1$ and $R_2$ are chosen, leading to $64$ combinations. If a QPSK symbol with radius $R_1$ is chosen for Pol X, an 8-PSK symbol with radius $R_2$ is selected for Pol Y. This configuration contains 32 distinct 4D symbols. Inversely, the other configuration also contains 32 symbols.

Now, we define the bit mapping rules to enhance the performance over an AWGN channel by maximizing the MSED and minimizing the bit error probability. The bit mapping is inspired by the one used in~\cite{ref1}. We decompose the 2D constellation in the I-Q plane into four areas defined by the four quadrants as shown in Fig.~\ref{fig:figure1} and we associate the symbols on the main axes in Pol X to the clockwise neighboring quadrant. Each 4D symbol is labeled by a six-bit vector $\left[ b_1,\dots,b_6\right]$ as follows:
\begin{itemize}
\item The quadrants in each polarization are defined by 4 out of the 6 bits, namely, $\left[b_1,b_2\right]$ and $\left[b_3,b_4\right]$ for the first and second polarization tributary respectively as we show in Fig.~\ref{fig:figure1}. These 4 bits determine which out of the 16 quadrant combinations is to be chosen.

\item  In each quadrant combination, there are 4 possible combinations of 2D symbols. $\left[b_5,b_6\right]$ are assigned to those as shown with the labels in Fig.~\ref{fig:figure1}.

\end{itemize}

\begin{figure}[htb]
    \centering
    \includegraphics[width=1\linewidth]{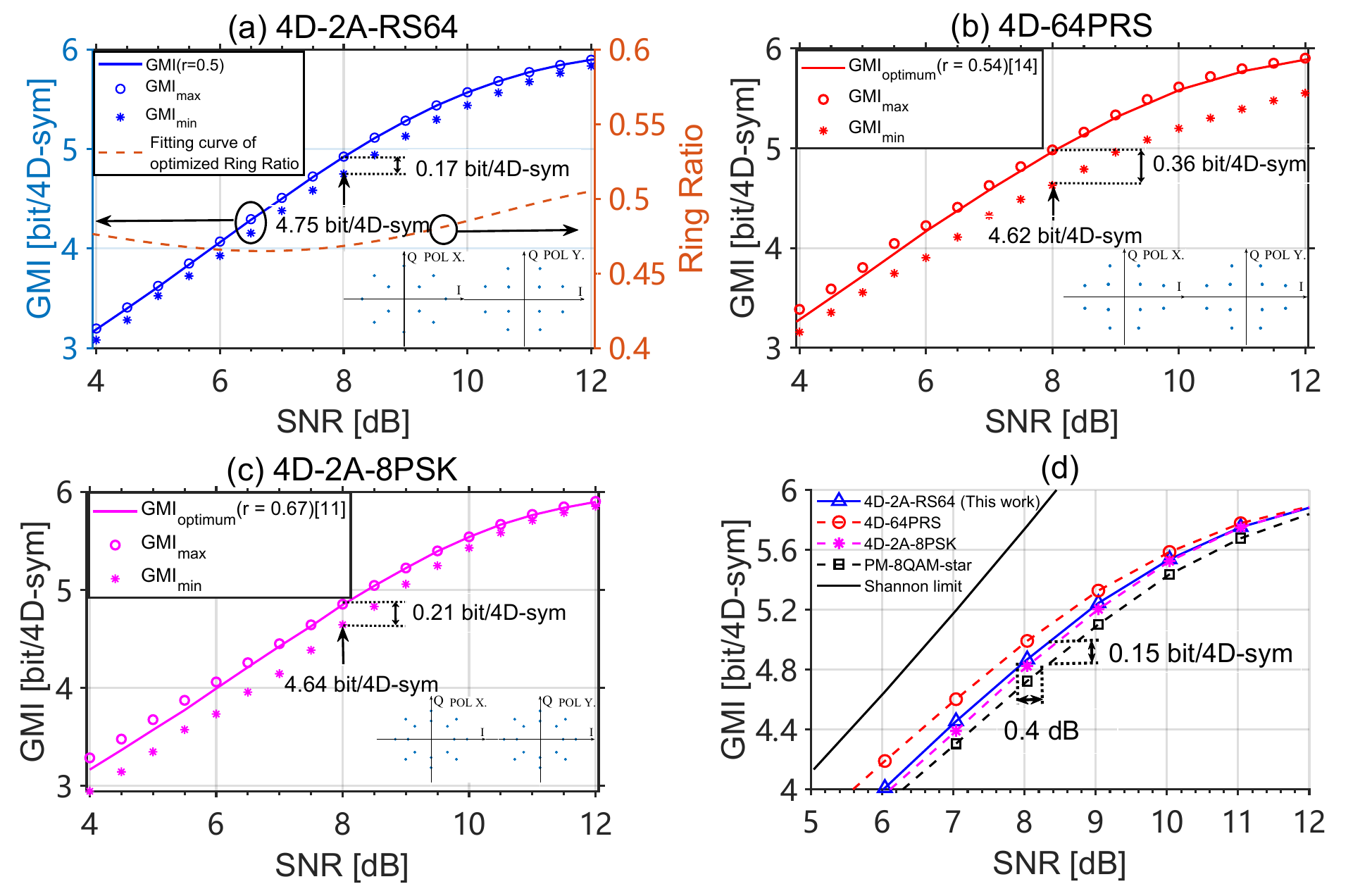}
    \caption{Solid line on (a), (b), (c): GMI (with optimized parameters for SNR=$8$~dB) versus SNR over AWGN channel; Dashed line on (a): Optimum ring ratio $r=R_1/R_2$ vs SNR. Asterisks: minimum GMI obtained by adding MZM imbalances of $\pm15^\circ$ for I-Q quadrature and $1.7$~dB for I-Q gain imbalances. Empty circles: maximum GMI when ring ratio is optimized for each SNR level. (d) GMI vs SNR for various modulation formats with 6 bit/4D-symbol. The insets within the figures show the 2D-projections of the three constellations.}
    \label{fig:GMI_R_Comparasion}
\end{figure}

A last optimization step of the constellation can typically be a geometric shaping consisting of changing the ring ratio between $R_1$ and $R_2$, $r=R_1/R_2$, or the relative orientation of the QPSK and 8-PSK constellations. Based on an optimization of the achievable information rates taking into account the bit mapping, the authors in \cite{ref1} and \cite{ref13} improved the linear performance of 4D-64PRS and 4D-2A-8PSK over a given signal-to-noise ratio (SNR) value for a target pre-FEC BER. However, a trade-off exists between the complexity of the implementation of the geometrically shaped constellation and the target performance. Indeed, overly precise optimization is difficult to achieve in practice because it increases the cost and the footprint of the transmitter (high effective number of bits (ENOB) requirements at high symbol rates). In Fig.~\ref{fig:GMI_R_Comparasion}(a), we report the optimal ring ratio (dashed-line curve and right y-axis) for each SNR and find it to be between $0.47$ and $0.51$. SNR is defined as the symbol energy over the power spectral density of the AWGN noise. In what follows, the ring ratio of our scheme is set to $0.5$, which is close to optimal over the studied SNR range.

In Fig.~\ref{fig:GMI_R_Comparasion}(d), we report the GMI of the three 4D schemes with optimized parameters at SNR$=8$~dB and no implementation penalties, and compare them to the performance of PDM-8QAM-star. At an FEC overhead of $\mathrm{25\%}$, corresponding to a net rate of $4.8$ bit/4D-symbol, we see that 4D-2A-RS64, 4D-64PRS and 4D-2A-8PSK have $0.4$ dB, $0.7$ dB and $0.3$ dB gain, respectively, over PDM-8QAM-star. The 4D-2A-RS64 performs better than 4D-2A-8PSK by $0.1$ dB. Conversely, at the SNR required by PDM-8QAM-star to achieve a net rate of $4.8$~bit/4D-symbols, 4D-2A-RS64 achieves a GMI gain of $0.15$~bit/4D-symbol compared to PDM-8QAM-star. 

\section{Tolerance to transmitter impairments}\label{Tolerance}

Before further investigating the performance of our proposed modulation, we test its robustness to ring ratio variations and relative rotation angle variations between the two rings and compare it to the robustness of 4D-64PRS and 4D-2A-8PSK. These ring and phase variations or deviations from their optimal values can be introduced by imbalances at the Mach-Zehnder modulator (MZM) such as gain imbalances, quadrature imbalances or even skew between I and Q components of each polarization tributary~\cite{Jacobs16,Liang21}. First, we fix the structure of the two-dimensional per-polarization projections of all three constellations using the optimized ring ratio for an SNR of $8$~dB. In this case, the performance of each 4D modulation is given by the full-line curves in Fig.~\ref{fig:GMI_R_Comparasion}(a-c) for SNR values between $4$ and $12$~dB. The optimal ring ratio value is also given in the legend of each sub-figure. The curves plotted with empty-circle markers show the maximum GMI values of each scheme when using the optimal ring ratio for each SNR value. Our proposed scheme shows a near-optimal performance with a fixed ring ratio in the studied SNR range. Later, we emulate gain and quadrature imbalances at the MZM using the following model~\cite{Jacobs16}:
\begin{equation}
\begin{split}
    E_x &= I_x +\alpha_x e^{j\theta_x}Q_x \\
    E_y &= I_y +\alpha_y e^{j\theta_y}Q_y
\end{split}
\end{equation}
where $I_x$ and $Q_x$ (resp. $I_y$ and $Q_y$) are the ideal in-phase and quadrature components of each polarization tributary, $E_x$ and $E_y$ are the generated complex amplitude at the outputs of each I-Q modulator, $\theta_x$ (resp. $\theta_y$) are the I-Q phase angles for the $x$ (resp. $y$) polarization tributary and $\alpha_x\leq1$ (resp. $\alpha_y\leq1$) is a gain discrepancy between the $I$ and the $Q$ components of each polarization tributary. When $\theta_x$ (resp. $\theta_y$) are equal to $90^\circ$, the quadrature is perfect between the I and Q components. When $\alpha_x$ and $\alpha_y$ are equal to 1, no gain disparity is introduced between the two components. According to an Optical Internetworking Forum (OIF) agreement on 400G transceivers~\cite{400ZR}, that is used as a basis for interoperability, the maximum allowable fluctuations for the MZM should be in the range of $\pm5^\circ$ for the I-Q quadrature imbalance and $1$~dB for the I-Q gain imbalance. To align with these requirements, these imbalances should be estimated and corrected through dedicated DSP algorithms and feedback loops. In our study, we chose the ranges for the considered imbalances similarly to previous works~\cite{Jacobs16,Fludger16,Liang21}, with worst-case values going beyond the typical values. In Fig.~\ref{fig:GMI_R_Comparasion}(a-c), the worst-case GMI values are shown through the curves with asterisk markers. These are obtained for maximum deviations of $\theta_x$ (resp. $\theta_y$) of $90^\circ\pm15^\circ$ and a maximum gain imbalance of $1.7$~dB between $I_x$ and $Q_x$ (resp. $I_y$ and $Q_y$) components. With these simulated imbalances, at SNR$=8$~dB, our 4D-2A-RS64 scheme has a minimum GMI floor of $4.75$~bit/4D-symbol and GMI fluctuation of $0.17$~bit/4D-sym. It shows better fault tolerance than 4D-64PRS and 4D-2A-8PSK, which have minimum GMI floors of $4.62$~bits/4D-sym and $4.64$~bits/4D-sym, and GMI fluctuations of $0.36$~bits/4D-sym and $0.21$~bits/4D-sym respectively.

\begin{figure}[t]
    \centering
    \includegraphics[width=1 \linewidth]{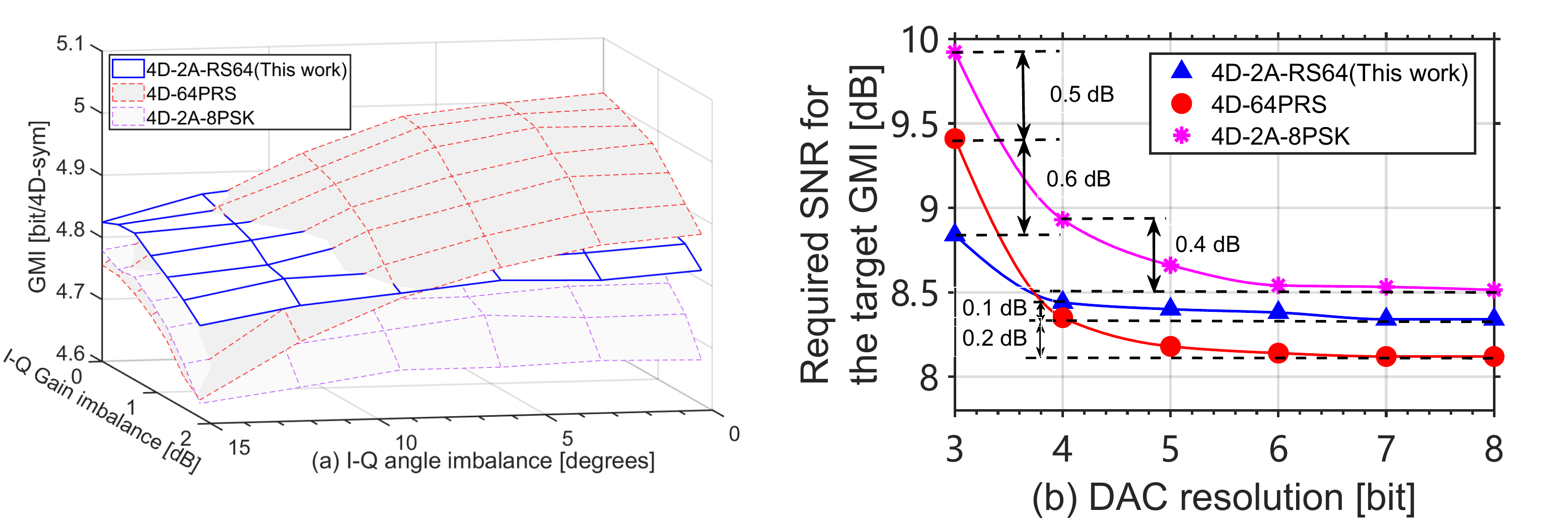}
    \caption{(a) GMI versus I-Q gain and quadrature imbalances at SNR=$8$~dB. (b) Required SNR versus DAC resolutions for different formats at a GMI of 5 bits/4D symbol.}
    \label{fig:MZM_Imbalance_and_DAC}
\end{figure}

To get a full view of the performance degradation of the three 4D schemes, we study the GMI variations for an I-Q gain imbalance within the range of $0$ to $1.8$~dB and an I-Q quadrature imbalance within the range of $0$ to $15$ degrees for each polarization tributary. In Fig.~\ref{fig:MZM_Imbalance_and_DAC}(a), we can see that, as the I-Q gain imbalance increases, the performance of 4D-64PRS deteriorates faster than the one of the proposed 4D-2A-RS64. The former becomes worse than the latter for I-Q angle imbalances higher than $12$~degrees in the absence of an I-Q gain imbalance, and higher than $8$~degrees for an I-Q gain imbalance of $1.8$~dB. It is important to note that all three 4D modulation schemes possess symmetrical characteristics along the I and Q axes; consequently, we obtain similar performance degradation for I-Q gain imbalances between $0$ to $-1.8$ dB and I-Q quadrature imbalances between $0$ and $-15$ degrees. 4D-64PRS requires the generation of precise amplitude and phase values to achieve the geometric shaping gain as shown in Fig.~\ref{fig:GMI_R_Comparasion}(b). As the MZM imbalance increases, its performance gradually deteriorates until it falls below that of 4D-2A-RS64. 4D-2A-8PSK comprises two 8PSK constellations, hence the inner ring of 8-PSK necessitates a fine amplitude and phase mapping, leading to continuous performance degradation as the imbalances increase; moreover, its initial imbalance-free performance is lower than that of 4D-2A-RS64.

An ideal digital-to-analog converter (DAC) is used in~\cite{ref2} and~\cite{ref14}. However, it is interesting to take into account the quantization error stemming from the finite resolution of DACs when evaluating the performance of geometric-shaped (GS) constellations. Moreover, high-resolution DACs not only add higher hardware costs but also a higher energy consumption~\cite{pfau2009hardware,almonacil2020performance}. To test the impact of DAC quantization, each transmitted sequence is oversampled to 2 samples per symbol, shaped using a root-raised cosine (RRC) filter with a roll-off factor equal to $0.1$, and uniformly quantized by a DAC with a vertical resolution between $3$ and $8$~bits. Even though the widely used DACs and ADCs in high-speed optical transceivers use $6$ to $8$ physical bits, the effective number of bits (ENoB) that take into account the signal distortions and noise added by the converters may drop below $4$~bits~\cite{Laperle13,Chen16}. In our study, we consider a frequency independent quantization DAC. The back-to-back transmission performance is measured over an AWGN channel at a symbol rate of $90$~Gbaud. In Fig.~\ref{fig:MZM_Imbalance_and_DAC}(b), we show the minimum required SNR to achieve a GMI of $5$~bit/4D-sym. For DAC resolutions ranging from 5 to 8 bits, there is a fine mapping for each constellation symbol and the SNR penalties are practically negligible. In such cases, as illustrated in Fig.~\ref{fig:GMI_R_Comparasion}(d), 4D-2A-RS64 incurs a slight SNR penalty compared to 4D-64PRS. At $4$~bits, the GMI penalties are $0.1$~dB, $0.2$~dB, and $0.4$~dB for 4D-2A-RS64, 4D-64PRS, and 4D-2A-8PSK respectively. In the meanwhile, 4D-64PRS requires the lowest SNR value. At $3$~bits, our 4D-2A-RS64 scheme shows a gain of $0.6$~dB and $~1.1$dB relative to 4D-64PRS and 4D-2A-8PSK respectively. The penalty of the 4D-64PRS comes from its sensitivity to phase changes; indeed, the phases of the outer ring constellation points cannot be accurately achieved with 3-bit resolution DACs. The 4D-2A-8PSK penalty comes from the use of an 8-PSK on its inner ring which requires a higher resolution to generate the constellation accurately. Although our proposed scheme outperforms the 4D-64PRS only when the studied transceiver impairments are severe (I-Q imbalance of more than $10$~degrees or a DAC resolution lower than $4$~bits), it still shows an advantage with respect to 4D-2A-8PSK which was implemented in commercially available transceivers as mentioned in~\cite{4DoverviewECOC23}. We believe that our proposed scheme merits to be considered as a relevant candidate among geometric-shaped (GS) 4D modulations as recent works are still looking into the design optimization of GS 4D schemes using machine learning tools~\cite{Yankov23} or simplified computation of AIR~\cite{Gumus23}.


\section{Performance evaluation}\label{Performance evaluation}
\subsection{Analysis of the performance over an AWGN channel}

\begin{table}[tb]\centering
    \renewcommand{\arraystretch}{1.2}
    \caption{ Metrics for different 4D constellations}
	\label{Tab 2}
\setlength{\tabcolsep}{0.7mm}
\begin{tabular}{|c|c|c|c|c|}
\hline
\textbf{}    & PDM-8QAM-star & 4D-2A-8PSK & 4D-64PRS & \textbf{4D-2A-RS64} \\ \hline
MSED         &  0.84    & 0.88   &  0.69    & \textbf{0.8}          \\ \hline
$n_{d}$           &  192    &  128  &   32   &    \textbf{64}       \\ \hline
\begin{tabular}[c]{@{}c@{}}$(d_{D_H=1}^{2}$,\\ $n_{D_H=1})$\end{tabular} & \begin{tabular}[c]{@{}c@{}}(0.84, 128)\\ (3.15, 64)\end{tabular} & \begin{tabular}[c]{@{}c@{}}(0.88, 128)\\ (3.46, 64)\end{tabular} & \begin{tabular}[c]{@{}c@{}}(0.69, 32)\\ (0.90, 64)\\ (0.98, 64)\\ (5.50, 32)\end{tabular} & \begin{tabular}[c]{@{}c@{}}\textbf{(0.8, 64)}\\ \textbf{(0.94, 64)}\\ \textbf{(1.39, 16)}\\ \textbf{(3.2, 32)}\\ \textbf{(5.46, 16)}\end{tabular} \\ \hline

Gray Labeled &  No    & Yes   &  Yes    &   \textbf{Yes}        \\ \hline
Constant Modulus           &  No    & Yes   &  Yes    &    \textbf{Yes}       \\ \hline
PAPR         &  1.58    &   1 &   1   &     \textbf{1}     \\ \hline
\end{tabular}
\end{table}

\begin{figure}[htb]
    \centering
    \includegraphics[width=1 \linewidth]{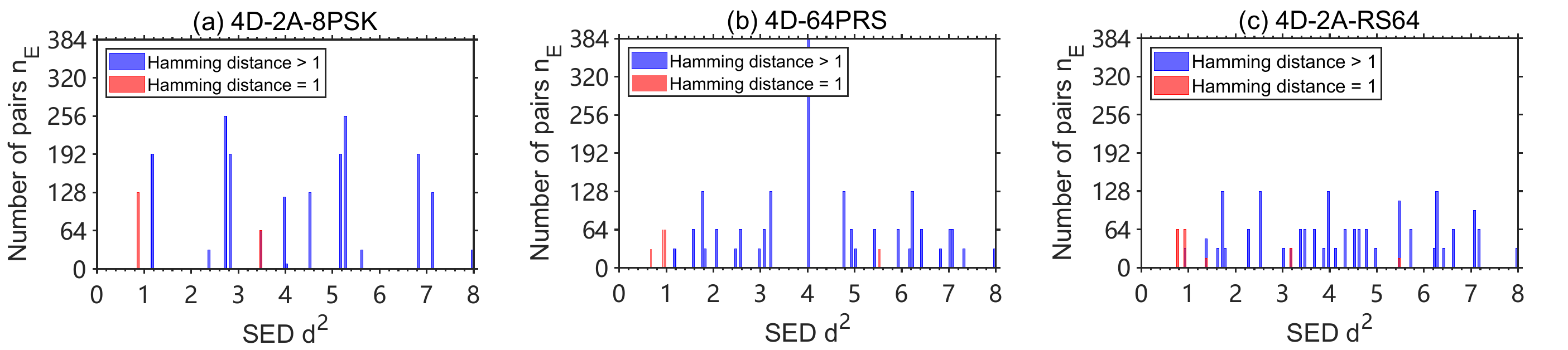}
    \caption{ Histograms of SEDs in three 4D formats: (a) 4D-2A8PSK, (b) 4D-64PRS, and (c) Our 4D-2A-RS64. The red bars show the number of pairs with Hamming distance of 1 at the SED $d^2$. The blue bars show the number of pairs with Hamming distance greater than 1 at the SED $d^2$.}
    \label{fig:4D-formats-MSED-HD}
\end{figure}


To easily compare the performance of the four formats, we list the following metrics and features for each format in Table \ref{Tab 2}:

\begin{itemize}
\item $\mathrm{MSED}$: the Minimum Squared Euclidean Distance between 4D-symbols. Each polarization tributary is normalized to a unit energy.
\item $n_{d}$: the number of pairs of 4D symbols at MSED. 
\item The pairs $(d_{D_H=1}^{2},n_{D_H=1})$ contain the values of Squared Euclidean Distances (SEDs) for which the Hamming distance ($D_H$) between the mapped bit patterns is equal to $1$ and the number of symbols at those SEDs.
\item Gray Labeled indicates whether all the constellation pairs at MSED are at a Hamming distance of 1. 
\item Constant Modulus indicates whether the 4D-energy is constant and PAPR is the peak-to-average power ratio. 
\end{itemize}

From Table~\ref{Tab 2}, we see that PDM-8QAM has the worst linear performance because it has a moderate MSED and an excessive $n_{d}$, in addition to the absence of Gray labeling. Among the three 4D formats, 4D-2A-8PSK has the largest MSED thus obtaining the best performance for high SNRs, followed by 4D-2A-RS64, then 4D-64PRS. In addition, among the four schemes, 4D-64PRS has the smallest number of symbol pairs at MSED, $n_{d}$, thus obtaining the best performance at low SNR values, followed by 4D-2A-RS64, then 4D-2A-8PSK with an increasing $n_d$. All three 4D formats are Gray-labeled and have a constant 4D modulus. Finally, to complement Table~\ref{Tab 2}, we show in Fig.~\ref{fig:4D-formats-MSED-HD} the SED distribution for the three 4D formats. From these histograms and the third row in Tab.~\ref{Tab 2}, we see that, although the 4D-64PRS has the smallest MSED, only $32$ pairs are located at this distance. Also, the pairs at the three smallest SEDs have a Hamming distance equal to $1$, which provides gains at low SNR values. The 4D-2A-RS64 constellation has a smaller MSED than the 4D-2A-8PSK, but only $64$ pairs are located at MSED compared to $128$ for the latter. In addition, for its $128$ pairs with the smallest SED, 4D-2A-RS64 has $64$ pairs at an SED of $0.94$ that contribute to a small performance improvement compared to 4D-2A-8PSK at low SNR values. The loss of performance compared to 4D-64PRS is mainly due to the $32$ symbol pairs at a SED of $0.94$ and with a Hamming distance larger than $1$ as can be seen in Fig.~\ref{fig:4D-formats-MSED-HD}(c).

\subsection{Performance over the non-linear channel}

Over long-haul multi-span fiber transmissions, NLI induced by the Kerr effect will add distortions to the propagating signal in addition to the additive ASE noise from the optical amplifiers. This NLI can be reduced by forcing a constant energy constraint on the transmitted signal, hence minimizing the peak-to-average power ratio (PAPR) at the transmitter side which leads to lower NLI generated by both intra-channel and inter-channel propagation effects~\cite{ref33}. We test the performance gain over fiber propagation through numerical simulations. We compare the BERs, system margins, and GMI values of the three 4D formats to the conventional format PDM-8QAM-star, all having an entropy of 6 bit/4D. The transmitted signal consists of $11$ WDM channels $\times~90$ Gbaud, each modulated over dual-polarization. The channel spacing is 100~GHz and the central channel is located at $1550$~nm. All channels are pulse-shaped with a root-raised-cosine filter (RRC) with a roll-off factor equal to $0.1$. In order to better compare linear gain, nonlinear gain, and gain changes under MZM imbalance, we used an ideal laser with no phase noise. The tested fibers are standard single mode fibers (SSMF) with a chromatic dispersion coefficient $D=17~\mathrm{ps/nm/km}$, a non-linearity coefficient $\gamma = 1.32~\mathrm{(W\cdot km)^{-1}}$, a fiber attenuation parameter $\alpha =0.2~\mathrm{dB/km}$. Polarization mode dispersion (PMD) is added with a PMD coefficient equal to $0.04~\mathrm{ ps/\sqrt{km}}$. The link configuration was a homogeneous multi-span link with $N\times80~\mathrm{km}$ spans. The propagation was modeled through a split-step Fourier method (SSMF) implementing the non-linear Manakov equation with 50 wave plates per span to simulate the polarization effects~\cite{ref4}. Each span was followed by an erbium-doped fiber amplifier (EDFA) with a noise figure of 5 dB. After propagation, the central channel is filtered and detected through a coherent receiver. We then apply on the detected signal a matched RRC filter with the same roll-off factor, and digitally compensate for the chromatic dispersion. Genie-aided linear channel equalization is then applied to compensate for the polarization changes prior to decoding (linear channel response is saved and fed to the frequency-domain MMSE channel equalizer). Finally, a genie-aided XPM-induced phase estimation filter with an averaging window of $\mathrm{\omega =64}$ symbols is used to compensate for the XPM-dominated phase rotation as in~\cite{ref5}. With these two perfect estimations, the measured performance is only related to the used modulation format (no penalty caused by imperfect channel estimation). No additional non-linear compensation algorithms such as digital back-propagation were used at the receiver side.

\begin{figure*}[htb]
    \centering
    \includegraphics[width=1 \linewidth]{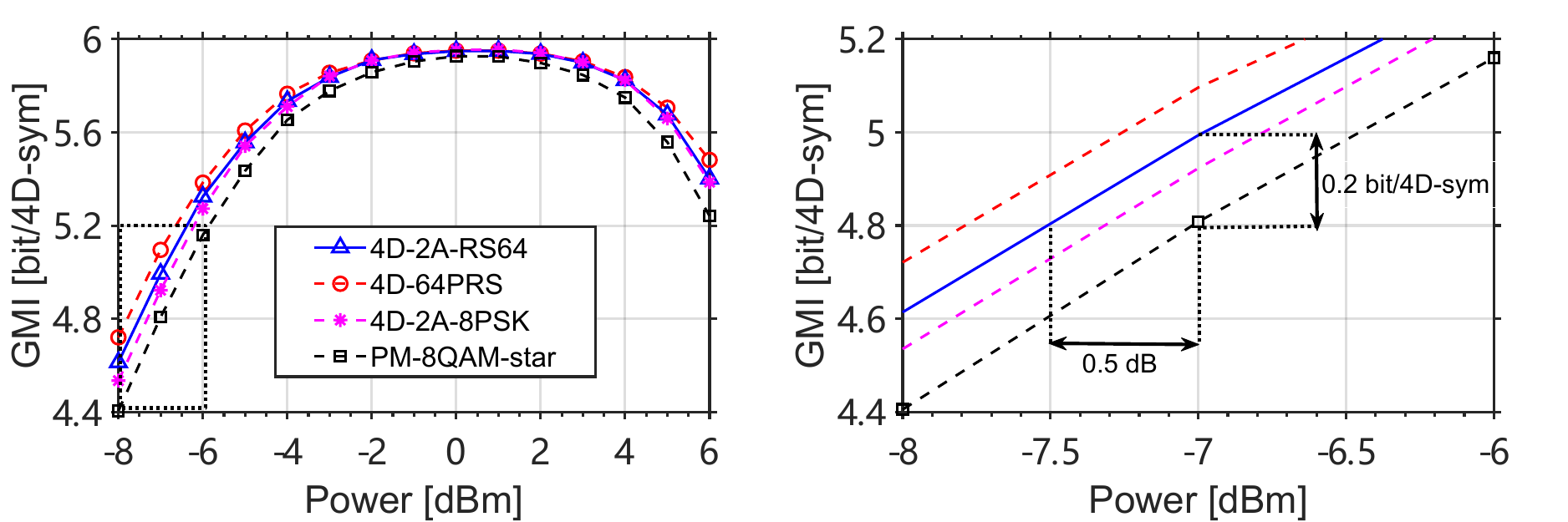}
    \caption{GMIs versus transmitted power after transmission over 20 spans of 80 km SSMF in the non-linear regime for 11 channels with 100 GHz spacing with a symbol rate of 90 Gbaud. Left: Overall trends. Right: Zoomed trends around a GMI of $4.8$~bit/4D.}
    \label{fig:AIR_Power_transmission}
\end{figure*}

Fig. \ref{fig:AIR_Power_transmission} shows GMI curves versus single-channel launch power values after transmission over 20 spans of 80km SSMF. All three 4D formats outperform PDM-8QAM-star in the linear regime as well as in the non-linear regime. At an FEC overhead of $\mathrm{25\%}$ corresponding to a net rate of $4.8$~bit/4D-symbol, we zoomed in on the corresponding area to better show the achieved gains. 4D-2A-RS64, 4D-64PRS and 4D-2A-8PSK show $0.5$ dB, $0.73$ dB and $0.25$ dB gains, respectively, over PDM-8QAM-star. 4D-2A-RS64 performs better than 4D-2A-8PSK by $0.25$ dB. While it does not beat the performance of 4D-64PRS, 4D-2A-RS64 does not require precise geometric shaping to maintain these gains and is less sensitive to implementation errors. At the SNR required for PDM-8QAM-star to reach 4.8 bit/4D-symbol, the GMI of 4D-2A-RS64 is 0.2bit/4D higher than PDM-8QAM-star. We obtain a higher gain in the fiber channel than in the back-to-back case because the 4D modulations are mitigating NLI in the presence of Kerr effect better than PDM-8QAM. It is also worth noting that in the single-channel test, we obtain nearly the same gains (not presented in Fig.~\ref{fig:AIR_Power_transmission}), which is attributed to the constant modulus characteristics of the three 4D formats that effectively reduce the generation of self-phase modulation (SPM).

\begin{figure}[htb]
    \centering
    \includegraphics[width=1 \linewidth]{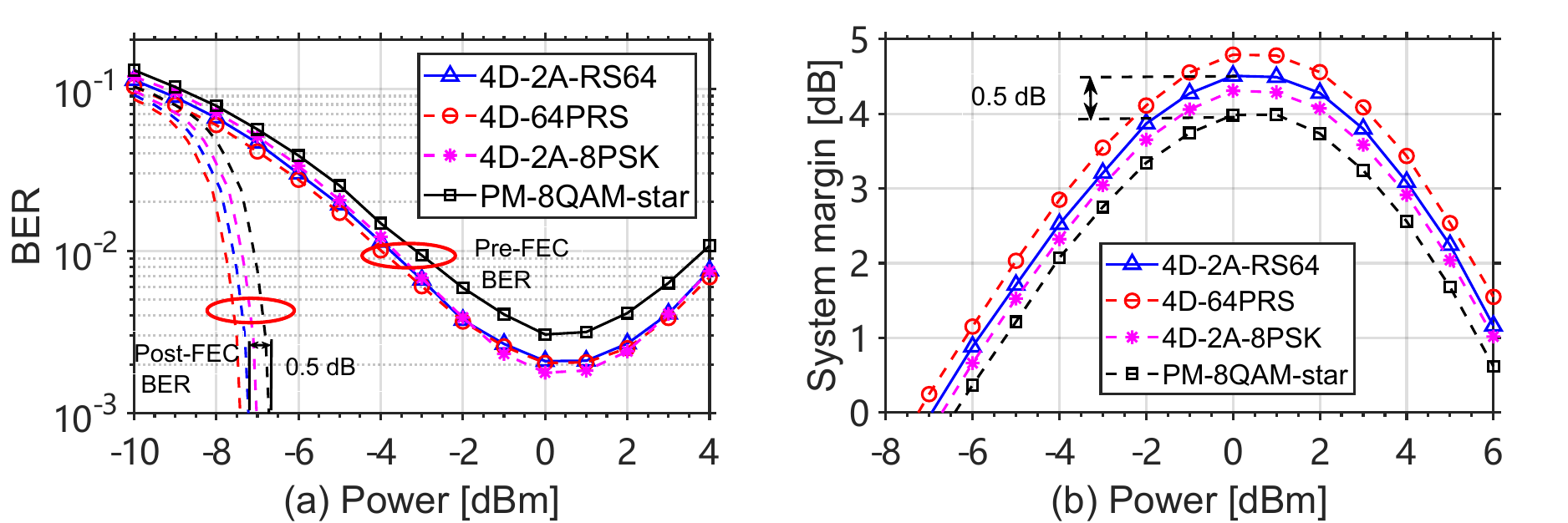}
    \caption{(a) Pre-FEC and post-FEC BERs versus transmitted power after transmission over 20 spans of 80km SSMF in the non-linear regime for 11 channels with 100 GHz spacing and a symbol rate of 90 Gbaud. (b) System margin (SNR gain compared to SNR with BER threshold = $4\times10^{-2}$) versus transmitted power after transmission over 20 spans of 80 km SSMF in the non-linear regime for 11 channels with 100 GHz spacing with a symbol rate of 90 Gbaud.}
    \label{fig:BER_Power_transmission_and_SNRmargin}
\end{figure}

In Fig. \ref{fig:BER_Power_transmission_and_SNRmargin}(a), we show post-FEC and pre-FEC BER curves versus launch power after transmission over 20 spans. We used low-density parity-check (LDPC) codewords of length $64800$~bits from the DVB-S2 standard~\cite{etsi2005digital} with a rate $R_C= 0.8$ and belief propagation (BP) decoding at the receiver side with 50 decoding iterations for a pre-FEC BER of $\mathrm{4\times 10^{-2}}$. The 4D-2A-RS64 gain in post-FEC BER compared to PDM-8QAM-star is in good agreement with the predictions of GMI gain of $0.5$~dB in Fig. \ref{fig:AIR_Power_transmission}. At the optimal power operation point, all three 4D formats have a lower pre-FEC BER than PDM-8QAM-star. 4D-2A-RS64 and 4D-64PRS achieve the same minimum pre-FEC BER while 4D-2A-8PSK has a slightly lower pre-FEC BER value. While the three 4D schemes have the same constant modulus property at the transmitter side, the lower minimum pre-FEC BER value of 4D-2A-8PSK is mainly due to its highest MSED value among the three schemes. 

As the GMI of the three 4D formats saturates around the optimal power in Fig.~\ref{fig:AIR_Power_transmission}, we also show in Fig.~\ref{fig:BER_Power_transmission_and_SNRmargin}(b) the system margin versus launch power for a target pre-FEC BER of $4\times10^{-2}$. The system margin is defined as the difference between the required $\mathrm{SNR_{elec}}$ to achieve a given target pre-FEC BER and the $\mathrm{SNR_{elec}}$ at any launch power of the transmission system over the studied link configuration. $\mathrm{SNR_{elec}}$ is the SNR measured from the constellation at the end of the DSP chain at the receiver side (i.e., at the input of the decision circuit). The system margin captures both the linear gain and the non-linear gain brought by the reduced energy variations of the considered 4D schemes. 4D-2A-RS64, 4D-64PRS, and 4D-2A-8PSK show $0.5$~dB, $0.8$~dB, and $0.3$~dB gain in system margin, respectively, over PDM-8QAM-star at the optimal launch power. Although chromatic dispersion will gradually increase the PAPR of the transmitted signal, we still measure gains when the signal has a constant-energy property at the transmitter side. We observe similar performance trends for the four schemes when propagating a single  $90~\mathrm{Gbaud}$ channel over the same link.

\begin{figure}[htb]
    \centering
    \includegraphics[width=1 \linewidth]{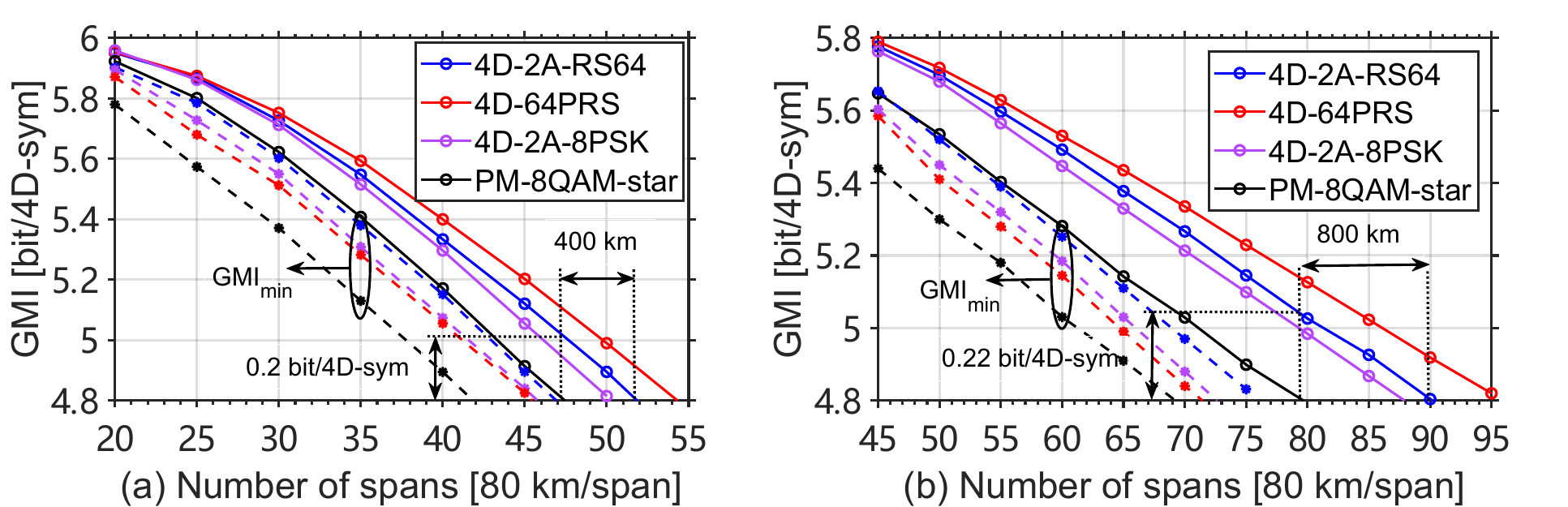}
    \caption{(a) GMIs versus the number of spans [80 km/span] for different 4D modulation formats with 6 bit/4D-symbol at 90 Gbaud. Solid lines with empty circles: GMI without MZM imbalances. Dashed line with filled circles:  minimum GMI obtained by adding MZM imbalances of $\pm15^\circ$ for the I-Q quadrature and $1.7$~dB for the I-Q gain imbalances. (b) GMIs versus the number of spans [80 km/span] for different 4D modulation formats with 6 bit/4D-symbol at 45 Gbaud.}
    \label{fig:AIR_NSPAN_transmission_90_45}
\end{figure}

Finally, the solid lines in Fig.~\ref{fig:AIR_NSPAN_transmission_90_45}(a) show GMI curves of the central channel versus the transmission distance given as a number of spans at the optimal launch power for each distance when transmitting $90~\mathrm{Gbaud}\times11$ channels. For a net rate of $4.8$ bit/4D-symbol, 4D-2A-RS64 increases the covered distance by $400$~km relative to PDM-8QAM-star and more than $160$~km relative to 4D-2A-8PSK, an increase of $\mathrm{10.6\%}$ and $\mathrm{4\%}$, respectively. 4D-64PRS obtains the best performance, allowing transmission over two extra spans compared to 4D-2A-RS64. At the maximum transmission distance of PDM-8QAM-star, which corresponds to 47 spans, 4D-2A-RS64 offers a GMI gain of $0.2$~bit/4D-symbol which is higher than the linear gain observed in Fig. \ref{fig:GMI_R_Comparasion}(d). The relative additional non-linear gain in terms of GMI for the nonlinear channel with respect to the gain in the linear channel is $(0.2 - 0.15)/0.15=33\%$, which is higher than the relative gains of 4D-2A-8PSK and 4D-64PRS, $(0.13 - 0.1)/0.1=30\%$ and $(0.29 - 0.26)/0.26=11.5\%$ respectively. Next, we add the same worst-case simulated MZM imbalances as the ones used for the linear AWGN channel. The dashed lines in Fig.~\ref{fig:AIR_NSPAN_transmission_90_45}(a) show that our 4D-2A-RS64 scheme has the best performance, allowing the transmission over almost 2 extra spans compared to 4D-64PRS and 5 extra spans compared to PDM-8QAM-star.

We also test the performance at a lower channel rate. In Fig.~\ref{fig:AIR_NSPAN_transmission_90_45}(b), we show the results for $45~\mathrm{Gbaud}\times11$ channels. First, for a perfect transmitter (solid lines), at a GMI of $4.8$~bit/4D-symbol, 4D-2A-RS64 increases the covered distance by 800km relative to PDM-8QAM-star and more than 160 km relative to 4D-2A-8PSK, an increase of $\mathrm{12.5\%}$ and $\mathrm{2.3\%}$ respectively. Since the dispersion-induced inter-symbol-interference grows more slowly over the propagation distance with symbols at $45$~Gbaud than at $90$~Gbaud, the PAPR has a slower increase over distance. Moreover, the optimal power will also be higher for higher baud rates, hence the non-linear effects will be magnified and will deteriorate the signal even more. If we compare gains in transmission distance for the three 4D constant modulus formats when moving from $45$~Gbaud to $90$~Gbaud, we notice that 4D-64PRS and 4D-2A-RS64 offer a gain of 8 spans (respectively 5) instead of 16 spans (respectively 10) with respect to PDM-8QAM, showing a gain reduction of $50\%$ while the gain of 4D-2A-8PSK reduces from 8 to 3 spans, showing a gain reduction of $62.5\%$. Finally, with the simulated worst-case MZM imbalances, our 4D-2A-RS64 also shows the best performance, allowing transmission over 5 extra spans compared to 4D-64PRS and 7 extra spans compared to PDM-8QAM-star.

\section{Conclusion}
\label{Conclusion}
We designed a new 4D constellation, named 4D-2A-RS64, with an entropy of $6$~bit/4D, which showed better linear performance and higher fiber non-linearity tolerance than conventional modulation formats PDM-8QAM-star or a recent 4D format 4D-2A-8PSK optimized for the non-linear channel. We demonstrated through numerical simulations that the maximum transmission distances are increased with respect to both formats for different symbol rates. While the proposed modulation did not beat the performance of another 4D competitor, the 4D-64PRS format, we demonstrated that 4D-2A-RS64 imposes less stringent constraints on the transmitter compared to the other 4D formats. Hence, its generation using non-ideal components (extreme imbalances in MZMs and limited-resolution DACs) shows smaller performance degradation than state-of-the-art 4D formats such as 4D-2A-8PSK and 4D-64PRS in both linear and non-linear transmission regimes. In the design of future high-order modulation schemes with higher spectral efficiency, we can still optimize linear performance by increasing MSED and considering the distribution of all squared Euclidean distances. For the nonlinear part, it is crucial to maintain temporal energy fluctuations as much as possible. Furthermore, higher spectral efficiency often requires more symbols and more precise mapping, which means that sensitivity to MZM imbalance and DAC variations will increase. Extending the proposed solution to achieve even higher spectral efficiency by using more rings remains an open problem for future exploration. In addition, the design of signaling schemes that are optimized for the non-linear fiber channel using both multi-dimensional modulations and probabilistic shaping could help in achieving additional linear and non-linear gains as it has been suggested in~\cite{Fehenberger16,Skvortcov21}.

\section*{Acknowledgments.}
This work was supported by Huawei Technologies France. We thank Djalal Bendimerad and Hartmut Hafermann for fruitful discussions on multi-dimensional formats.

\section*{Disclosures.}
The authors declare no conflict of interest.

\section*{Data availability.}
Data underlying the results presented in this paper are not publicly available at this time but may
be obtained from the authors upon reasonable request.


 


\bibliography{main}{}

\end{document}